\documentclass[a4paper,fleqn]{cas-dc}

\usepackage{graphicx}
\usepackage{algorithm}
\usepackage{algpseudocode}
\usepackage{xcolor}
\usepackage{listings}
\usepackage{amsfonts}
\usepackage{subcaption}

\usepackage[most]{tcolorbox}

\newtcolorbox{paperquote}{
  colback=gray!5,
  colframe=gray!40,
  boxrule=0.3pt,
  arc=1.5mm,
  left=6pt,right=6pt,top=4pt,bottom=4pt,
  fontupper=\itshape
}

\newcolumntype{L}[1]{>{\raggedright\let\newline\\\arraybackslash\hspace{0pt}}m{#1}}
\newcolumntype{C}[1]{>{\centering\let\newline\\\arraybackslash\hspace{0pt}}m{#1}}
\newcolumntype{R}[1]{>{\raggedleft\let\newline\\\arraybackslash\hspace{0pt}}m{#1}}
\definecolor{listinggray}{gray}{0.9}
\definecolor{lbcolor}{rgb}{0.9,0.9,0.9}
\usepackage{amsmath}
\usepackage[sort&compress,numbers]{natbib}
\usepackage[super]{nth}
\usepackage{balance}
\usepackage{xurl}

\usepackage{xcolor,colortbl}
\newcommand{\reviewSI}[1]{\textcolor{blue}{#1}} 

\usepackage{todonotes}

\begin{document}
\let\WriteBookmarks\relax
\def\floatpagepagefraction{1}
\def\textpagefraction{.001}
 
\shorttitle{Exploring the Role of LLMs in HPC Programming: A Survey}    
\shortauthors{S. Ljaljevic, J. Jorba and S. Iserte}  
\title [mode = title]{Exploring the Role of Large Language Models in High-performance Computing Programming: A Survey}

\author[1]{Strahinja Ljaljevic}[orcid=0009-0006-7003-4574]

\author[1]{Josep Jorba}[orcid=0000-0002-5810-4748]

\author[2]{Sergio Iserte}[orcid=0000-0003-3654-7924]
\ead{sergio.iserte@bsc.es}

\affiliation[1]{organization={Universitat Oberta de Catalunya (UOC)},
            city={Barcelona},
            country={Spain}}

\affiliation[2]{organization={Barcelona Supercomputing Center (BSC)}, 
            city={Barcelona},
            country={Spain}
            }

\begin{abstract}
Large Language Models (LLMs) are emerging as promising assistants in High-Performance Computing (HPC), where programming remains complex and expertise-intensive. This survey systematically reviews their application across five categories: code generation, parallelization and optimization, frameworks and architectures, evaluation and benchmarking, and broader challenges. The analysis highlights both opportunities and limitations: while general-purpose LLMs perform reasonably well on serial and OpenMP-like tasks, they fall short in distributed paradigms such as MPI, where correctness and scalability are critical. Domain-specialized models (e.g., HPC-Coder, HPC-GPT, chatHPC) achieve higher accuracy through fine-tuning, curated datasets, and retrieval-augmented generation (RAG), yet their scope remains narrow and their evaluations largely limited to benchmarks or micro-kernels.
The broader picture is one of dual potential and fragility: LLMs can lower barriers to entry, accelerate prototyping, and support code modernization, but they remain brittle under production-level requirements where correctness, performance portability, and scaling cannot be compromised.

We conclude that LLMs are unlikely to replace HPC experts in the near term but are positioned to become powerful collaborators in the software development pipeline. Their effective deployment will require richer datasets, integration with performance analysis and schedulers, rigorous evaluation frameworks, and governance structures that ensure transparency and trust. The convergence of AI and HPC should therefore be understood as a long-term, co-evolutionary process—where each advance uncovers new challenges and opportunities for reshaping scientific software development.
\end{abstract}

\begin{keywords}
HPC \sep 
Parallel Programming \sep
Domain-Specific LLMs \sep
State of the Art \sep
AI for Scientific Computing
\end{keywords}

\maketitle

\section{Introduction}\label{intro}
This survey explores how Large Language Models (LLMs), originally designed for natural language processing tasks, are being leveraged to support High-Performance Computing (HPC) workflows. As parallel programming grows increasingly complex and specialized, the potential for LLMs to act as intelligent assistants in code generation, performance optimization, and system-level guidance has sparked growing research interest. This work aims to examine this intersection by critically analyzing current research efforts and identifying where these models succeed, where they fall short, and what opportunities lie ahead.

LLMs are particularly promising for HPC for several reasons. First, they can bridge expertise gaps by providing contextual assistance to developers who may lack deep experience in advanced programming paradigms, thus lowering the entry barrier for new researchers in the field. Second, they can accelerate development by automating boilerplate and repetitive coding tasks, enabling developers to focus on algorithmic design and performance-critical decisions. Third, they facilitate exploratory workflows, allowing scientists and engineers to rapidly prototype new programming models, computational strategies, or performance optimizations without committing to extensive manual code refactoring. Finally, LLMs can assist hybrid heterogeneous parallel programming models by suggesting platform-specific optimizations for diverse architectures (e.g., OpenMP, MPI, CUDA). In practice, these capabilities translate to tangible benefits such as generating hybrid kernels, adapting job scripts for varying cluster configurations, or proposing transformations that improve memory locality and scalability.

To build a strong foundation, this work outlines the technical challenges of HPC programming, ranging from shared and distributed memory paradigms to heterogeneous frameworks, and by reviewing the evolution of LLMs into systems capable of producing high-level and low-level code. A comprehensive analysis of state-of-the-art research is then presented, organized into five thematic categories: 1) Code Generation, 2) Parallelization and Optimization, 3) Frameworks and Architectures, 4) Evaluation and Benchmarking, and 5) Challenges and Landscape. Our findings highlight both promising capabilities---such as LLMs generating OpenMP kernels or assisting in loop transformations---and persistent limitations as we have observed in token size constraints, inconsistent output quality, and the absence of HPC-specific benchmarks.

To ensure a comprehensive and relevant overview, the survey applies a structured literature screening methodology. Sources were identified through systematic searches in academic databases such as IEEE Xplore, ACM Digital Library, arXiv, and SpringerLink, using targeted keywords including \textit{LLM for HPC, parallel programming} and \textit{code generation}. Articles were selected based on their relevance to LLM-HPC integration, technical depth and recency (with a focus on publications from 2023-onward). Peer-reviewed papers and high-quality preprints were prioritized, while non-technical, non-peer-reviewed or general-purpose AI studies without HPC context were excluded.

The survey identifies gaps in training data, model design, and usability, and proposes concrete directions for future research, such as integrating LLMs with Graph Neural Networks (GNNs) and developing interactive coding environments tailored for parallel programming and HPC-specific needs. By offering a structured synthesis of this emerging field, this work encourages deeper collaboration between the AI and HPC communities to responsibly and effectively integrate LLMs into scientific computing workflows.

The rest of the paper is structured as follows: Section~\ref{sec:background} introduces the background in LLMs. Section~\ref{sec:methodology} describes the methodology followed to identify sources and presents the criteria for inclusion or exclusion of cited articles, and exposes the research questions that motivate this study. Section~\ref{sec:sota} critically analyzes the included references and scrutinizes their supporting data. Finally, Section~\ref{sec:conclusions} concludes the paper by answering the raised research questions, discussing the ethical impact, and proposing a series of future research directions.

\section{Background}\label{sec:background}
This section provides context on the intended use of LLMs, their historical development, and their application in the domain of computer programming, particularly in HPC.

\subsection{The Purpose of an LLM}
LLMs are deep learning models trained to understand and generate human-like language. Built predominantly on the Transformer architecture~\cite{vaswani_attention_2017}, they are capable of performing a variety of tasks such as question answering, text summarization, translation, and, increasingly, compute code generation. The core purpose of an LLM is to model the probability distribution over sequences of words (or tokens) in a way that captures syntax, semantics and context. In doing so, they can generate coherent outputs that mimic natural language structures.

For code-related tasks, this modeling capability is extended to programming languages. LLMs trained on large codebases acquire the ability to autocomplete functions, generate code from natural language prompts, identify and correct bugs and even suggest performance improvements.

\subsection{Evolution of LLMs}
The evolution of LLMs follows a trajectory from symbolic rule-based systems to today's multi-billion parameter~\cite{touvron_llama_2023,brown_language_2020} deep learning models. Early efforts in natural language modeling relied on statistical techniques like N-grams~\cite{jurafsky_speech_2025}, which were limited in their ability to capture long-range dependencies. This changed with the emergence of neural networks and, subsequently, the introduction of the Transformer architecture~\cite{vaswani_attention_2017}.

Transformers enabled models to attend to all parts of an input sequence simultaneously, a property that significantly improved performance in tasks requiring contextual understanding. OpenAI's GPT series\footnote{\url{https://chatgpt.com}}, Google's BERT\footnote{\url{https://github.com/google-research/bert}}, and Meta's LLaMA\footnote{\url{https://www.llama.com}} are prominent milestones in this evolution. These models have grown in size and capability, progressing from hundreds of millions to trillions of parameters and capable of few-shot and zero-shot learning. Recent iterations, such as GPT-5 and Claude, incorporate Reinforcement Learning from Human Feedback (RLHF) to enhance alignment with user expectations, though this technique can also introduce side effects, such as models being overly cautious or "polite" in their response.

\subsection{LLMs for Computer Programming}

Code-specific large language models, often referred to as Code LLMs, extend the general language modeling paradigm to the domain of computer programming. These models are trained on extensive datasets comprising source code written in multiple programming languages (e.g., C, C++, Python, Fortran) as well as related documentation and natural language descriptions. This dual training enables them to generate code, interpret and comment on existing code, and bridge the gap between natural and programming languages.

The first widely known breakthrough in this field was Codex~\cite{chen_evaluating_2021}, which powered GitHub Copilot and demonstrated the feasibility of using LLMs for general-purpose code generation. Later, AlphaCode~\cite{li_competition-level_2022} explored competitive programming tasks, showcasing that LLMs could reason about complex algorithmic problems. More recently, open-source models like Code Llama~\cite{roziere_code_2024}, StarCoder~\cite{li_starcoder_2023}, and DeepSeek-Coder~\cite{guo_deepseek-coder_2024} have extended these capabilities with broader language support, improved training pipelines, and permissive licenses, making them particularly relevant for research environments. Table~\ref{tab:code-llms} summarizes these Code LLMs. 

\begin{table*}[tbp]
\centering
\fontsize{9.5pt}{14pt}\selectfont
\caption{Representative Code LLMs relevant to HPC programming.}
\label{tab:code-llms}
\begin{tabular}{l c c c l}
\toprule
\textbf{Model} & \textbf{Year} & \textbf{Parameters} & \textbf{License} & \textbf{Notable Features} \\ \midrule
Codex~\cite{chen_evaluating_2021} & 2021 & $\sim$12B & Proprietary & Powering GitHub Copilot, multi-language support. \\ 
AlphaCode~\cite{li_competition-level_2022} & 2022 & $\sim$41B & Proprietary & Competitive programming and problem-solving. \\ 
Code Llama~\cite{roziere_code_2024} & 2023 & 7B--34B & Open-source & Multi-language code generation, fine-tuned on code. \\ 
StarCoder~\cite{li_starcoder_2023} & 2023 & 15B & Open-source & BigCode dataset, permissive license for research. \\ 
DeepSeek-Coder~\cite{guo_deepseek-coder_2024} & 2024 & 7B--33B & Open-source & High-quality code completions, multi-language focus. \\ 
\bottomrule
\end{tabular}
\end{table*}

\subsection{LLMs for Parallel Programming}
Within the context of scientific computing and HPC, LLMs hold the potential to support developers throughout the software development lifecycle. Their applications include:

\begin{itemize}
\item \textbf{Code generation}: Writing parallel code for MPI, OpenMP, CUDA, or other parallel programming models.
\item \textbf{Code optimization}: Proposing improvements for memory usage, vectorization, and communication patterns.
\item \textbf{Debugging and refactoring}: Detecting race conditions, deadlocks, and inefficient constructs, and suggesting refactored alternatives.
\item \textbf{Documentation and education}: Explaining complex HPC algorithms, kernels, or parallelization patterns in an accessible manner.
\item \textbf{Porting and modernization}: Assisting in migrating legacy Fortran/C applications to contemporary programming models and heterogeneous hardware (e.g., GPUs, FPGAs, QPUs).
\end{itemize}

Several practical integration pathways have been explored to leverage these capabilities in HPC workflows, such as embedding LLMs in interactive notebooks to provide in-context code snippets, developing IDE plugins or leveraging existing LLM-based assistants that assist in writing MPI/OpenMP/CUDA code \cite{godoy_large_2024}, and deploying chatbots within HPC center portals to help users generate job scripts or configure software environments \cite{yin_chathpc_2024}. More advanced approaches include automated tuning frameworks, where LLMs generate multiple code variants for benchmarking and performance comparison \cite{wei_improving_2025}.

An example of a practical LLM prompt in this context could be:
\textit{
``Generate a hybrid MPI+OpenMP loop to perform a 3D stencil computation, ensuring minimal false sharing on NUMA architectures.''
}

\section{Methodology}\label{sec:methodology}
This survey employs a structured review methodology to systematically synthesize current research at the intersection of LLMs and HPC. The primary objective is to map the state of the art and derive a set of research questions that can guide future investigations.

\subsection{Corpora Compilation}
To ensure transparency, reproducibility, and extendibility, the corpora\footnote{\url{https://github.com/sljaljevic/llm-hpc-survey}} of reviewed works was compiled following a well-defined search strategy. This strategy combined multiple academic databases, a curated set of keywords, and explicit inclusion and exclusion criteria. The selected literature was subsequently categorized and analyzed within a thematic framework, facilitating both replication of the review and its extension with new publications.

The review focuses on peer-reviewed works published from 2023 onward, when the first dedicated studies on LLMs for HPC began to emerge. Earlier works predominantly addressed general-purpose NLP or preliminary explorations of code generation without explicitly targeting HPC-specific challenges such as parallel programming, communication patterns, or heterogeneous architectures.

Searches were performed across the following sources:
\begin{itemize}
    \item Google Scholar\footnote{\url{https://scholar.google.com}},
    \item IEEE Xplore\footnote{\url{https://ieeexplore.ieee.org/Xplore}},
    \item ACM Digital Library\footnote{\url{https://dl.acm.org}},
    \item arXiv\footnote{\url{https://arxiv.org}}, and
    \item SpringerLink\footnote{\url{https://link.springer.com}}.
\end{itemize}

The keyword set combined general and domain-specific terminology, including:
\begin{itemize}
    \item \textit{Large Language Models}, \textit{LLM},
    \item \textit{High Performance Computing}, \textit{HPC},
    \item \textit{Parallel Programming},
    \item \textit{Code Generation}, \textit{AI-assisted Programming},
    \item \textit{LLM for HPC}, \textit{LLM for Parallel Programming},
    \item \textit{Parallel Code Generation}, and
    \item \textit{Parallel Code Automation}.
\end{itemize}

It is important to note that the number of retrieved results varied significantly depending on the phrasing of keywords. For example, acronyms (LLM, HPC) often returned fewer results than their expanded forms (\textit{Large Language Models}, \textit{High Performance Computing}), especially when combined with additional descriptors. This observation highlights the importance of careful keyword selection when conducting systematic literature reviews or bibliometric analyses.

Each selected paper was then examined according to: 
\begin{enumerate}
    \item \textbf{The type of LLM employed:} Whether the study used a general-purpose model (e.g., GPT-3.5, GPT-4, LLaMA-2) or a code-specific model (e.g., StarCoder, CodeLlama, HPC-Coder).
    \item \textbf{Targeted programming models:} Covering shared-memory (OpenMP), distributed-memory (MPI), GPU programming (CUDA, HIP, OpenCL, SYCL), and task-based or hybrid approaches.
    \item \textbf{Evaluation strategies applied:} Including syntactic and semantic correctness, compilation success rates, runtime performance, speedup and efficiency metrics, scaling behavior, integration, and unit test results and human expert validation.
    \item \textbf{Depth of integration into HPC workflows:} Ranging from IDE-integrated assistants~\cite{godoy_large_2024}, to standalone frameworks (such as P4OMP~\cite{abdullah_p4omp_2025} or AutoParLLM~\cite{mahmud_autoparllm_2025}), to conversational agents and portals (like ChatHPC~\cite{yin_chathpc_2024} or HyCE~\cite{miyashita_llm_2024}) and full multi-agent systems for library generation (ChatBLAS~\cite{valero-lara_chatblas_2024}).
\end{enumerate}

Tables~\ref{tab:inclusion_criteria} and \ref{tab:exclusion_criteria} summarize the inclusion and exclusion criteria applied in this survey.

\begin{table*}[tbp]
    \centering
    \fontsize{9.5pt}{14pt}\selectfont
    \caption{Inclusion Criteria for Selected Literature}
    \label{tab:inclusion_criteria}
    \begin{tabular}{l p{11cm}}
    \toprule
    \textbf{Inclusion Criteria} & \textbf{Justification} \\
    \midrule
    Focus on LLM-HPC integration & The work must explicitly address the use of LLMs to support HPC workflows (e.g., code generation, parallelization, performance tuning). \\
    Peer-reviewed or recent preprints & Priority was given to peer-reviewed publications from 2023 onward; selected arXiv preprints were included for recency and relevance. \\
    Technical depth and evaluation & Works must include benchmarks, implementation details, or case studies that substantiate their claims. \\
    \bottomrule
    \end{tabular}
\end{table*}

\begin{table*}[tbp]
    \centering
    \fontsize{9.5pt}{14pt}\selectfont
    \caption{Exclusion Criteria for Selected Literature}
    \label{tab:exclusion_criteria}
    \begin{tabular}{l p{11cm}}
    \toprule
    \textbf{Exclusion Criteria} & \textbf{Justification} \\
    \midrule
    General AI papers & Works not addressing HPC or parallel computing were excluded, unless cited for foundational background. \\
    Non-technical or anecdotal articles & Blog posts, opinion pieces, or works lacking empirical data and reproducibility were not considered. \\
    Papers with significant overlap & Duplicates or derivative works without additional insights were excluded in favor of more comprehensive studies. \\
    \bottomrule
    \end{tabular}
\end{table*}

\subsection{Research Questions}
With this survey, we aim to give answers to the following research questions (RQ). These research questions are designed to guide a structured exploration of the field, enabling a critical synthesis of current research efforts, identifying where LLMs provide value in HPC and highlighting limitations, methodological gaps, as well as future opportunities. They ensure the review is not only descriptive but also analytical, providing clear insights for both researchers and practitioners.
\begin{itemize}
    \item RQ1: In which ways have LLMs been utilized in HPC domain so far?
    \item RQ2: Based on current research, what are the primary techniques and modes used to leverage LLMs for HPC?
    \item RQ3: How effective are LLMs in assisting with parallel programming and HPC-related tasks?
    \item RQ4: What evaluation metrics and benchmarks are used to assess LLM-generated code for HPC?
    \item RQ5: How do LLM-generated HPC solutions compare to human-written code in terms of correctness, efficiency, and scalability?
    \item RQ6: What are the key challenges and limitations of using LLMs for HPC?
    \item RQ7: What research gaps exist in the application of LLMs to HPC?
\end{itemize}

\section{State of the Art: LLMs for HPC Programming}\label{sec:sota}

This section describes, analyses, and correlates research on the use of LLMs for HPC, divided into five thematic areas: Code Generation (Subsection~\ref{subsec:code}), Parallelization and Optimization (Subsection~\ref{subsec:parallel}), Evaluation and Benchmarking (Subsection~\ref{subsec:evaluation}), Frameworks and Architectures (Subsection~\ref{subsec:framework}), and Challenges and Landscape (Subsection~\ref{subsec:challenges}). 

\begin{figure}[tbp]
    \centering
    \includegraphics[clip,width=0.85\columnwidth,trim={2.25cm 0.75cm 2.75cm 0.25cm}]{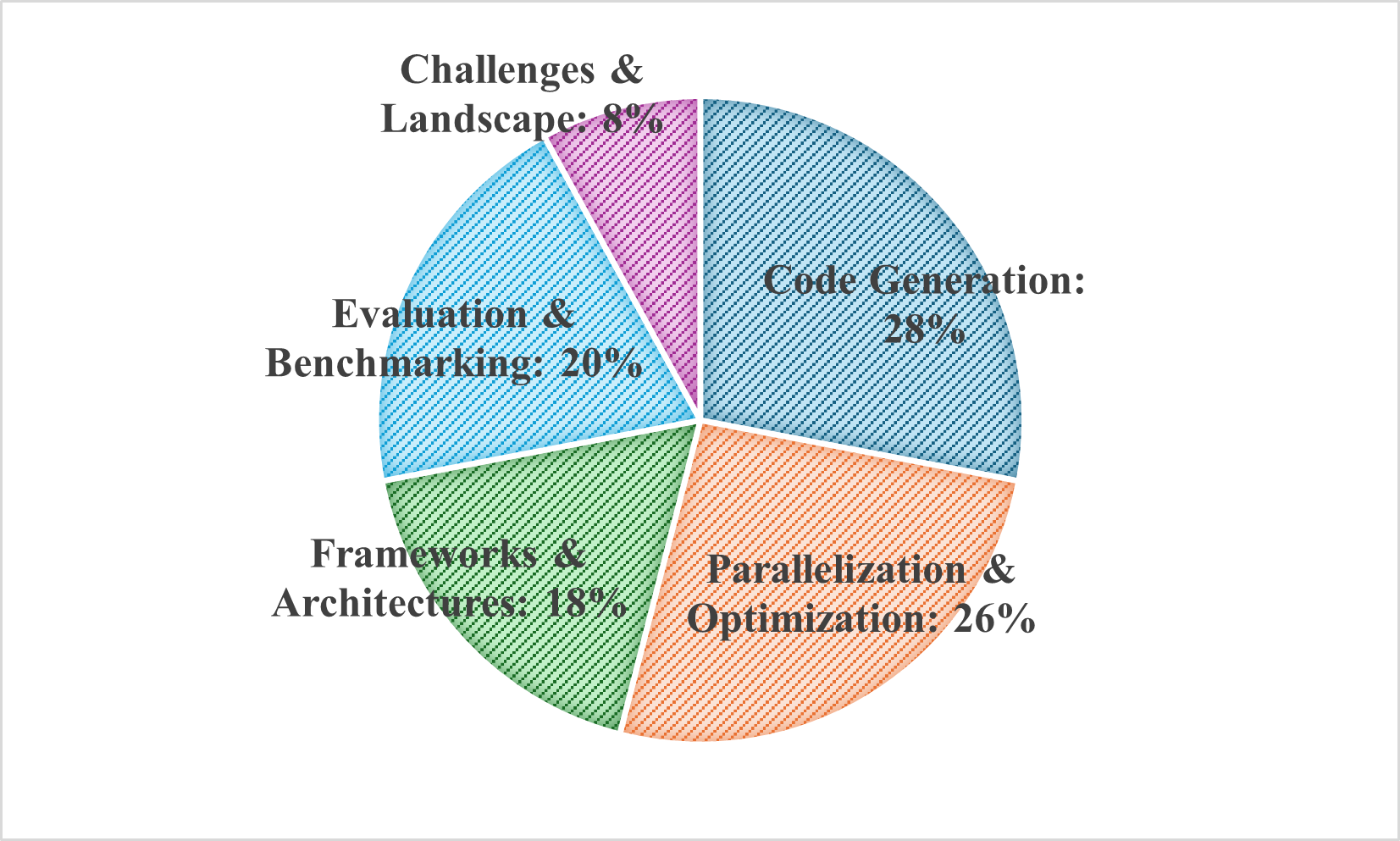}
    \caption{\reviewSI{Research papers categorized by LLM-HPC topic researched.}}
    \label{fig:papers_categorized}
\end{figure}

\begin{figure}[tbp]
    \centering
    \includegraphics[clip,width=0.6\columnwidth,trim={0.25cm 0.25cm 0.25cm 1cm}]{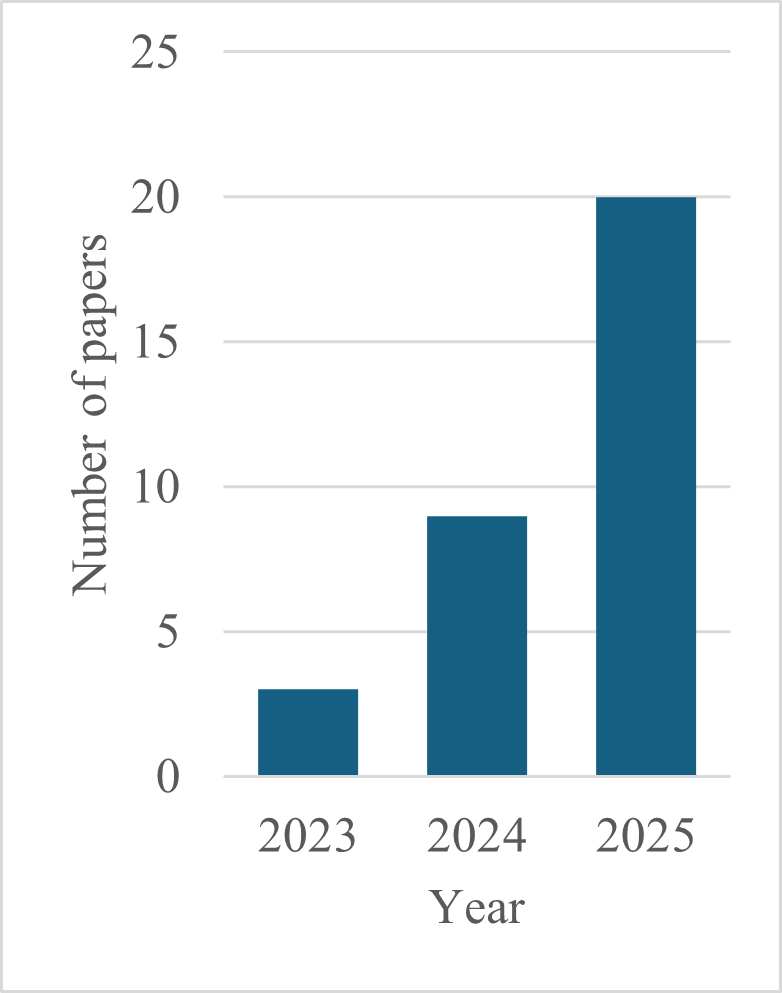}
    \caption{\reviewSI{Number of LLM–HPC papers published per year.}}
    \label{fig:papers_per_year}
\end{figure}

In total, \reviewSI{32 scientific articles out of 72} (see \textit{Data Availability} Section), met the inclusion criteria defined in the previous subsection. The distribution of these works across the five thematic categories is shown in \autoref{fig:papers_categorized}. Furthermore, \autoref{fig:papers_per_year} illustrates the upward trend in this emerging research area, highlighting both the increasing density and the growing number of LLM--HPC publications over time, \reviewSI{organized by year.}

To assess the performance of LLMs in HPC programming, different metrics are employed across the literature. These metrics capture aspects ranging from functional correctness to runtime efficiency, reflecting both AI-oriented and HPC-oriented perspectives. In this survey, we consider the most frequently reported measures: \texttt{pass@1}, \texttt{pass@k}, speedup, and efficiency. Their definitions and relevance are summarized in Table~\ref{tab:evaluation_metrics}.
Particularly, Nichols et al.~\cite{nichols_can_2024} (in \textit{ParEval} paper) as well as Davis et al.~\cite{davis_pareval-repo_2025} et al. (in \textit{ParEval-Repo} paper) use these as primary metrics. The main difference is that while \texttt{pass@1} denotes the fraction of tasks solved correctly by the model’s first output (top-1 sample), \texttt{pass@k} denotes whether any of the top-k sampled outputs is correct. In other words, \texttt{pass@1} is sensitive to initial prompt quality and model calibration; \texttt{pass@k} additionally reflects the model’s diversity under sampling. However, \texttt{pass@k} does not imply interactive prompting---it actually aggregates multiple independent generations from a fixed prompt.

\begin{table*}[tbp]
\centering
\caption{Key evaluation metrics used to assess LLMs in HPC programming.}
\label{tab:evaluation_metrics}
\begin{tabular}{r|p{13cm}}
\toprule
\textbf{Metric} & \textbf{Description} \\ \midrule
\texttt{pass@1} & Probability that the first generated code solution is correct. A strict and realistic measure since it assumes no resampling. \\ \midrule
\texttt{pass@k} & Probability that at least one correct solution exists among the top-$k$ generated outputs. Highlights improvements from multiple sampling.\\ \midrule
Speedup & Ratio of execution time of baseline implementation to that of the LLM-generated code. Used to evaluate performance gains (e.g., $2.3\times$ faster).\\ \midrule
Efficiency & Ratio between the generated code speedup and the ideal linear speedup relative to the number of processing elements. 
\\\bottomrule
\end{tabular}
\end{table*}

\subsection{Code Generation}\label{subsec:code}
The application of both general-purpose and fine-tuned Code LLMs in generating parallel code has gained significant traction in recent years. These models are leveraged for translating natural language descriptions into executable parallel code and for supporting domain-specific HPC workflows. Research in this category focuses on syntactic correctness, semantic fidelity and the effectiveness of prompt engineering techniques in guiding model outputs for HPC tasks. Studies reveal varied success depending on the model architecture, training datasets and the targeted parallel programming paradigms. 

Valero-Lara et al.~\cite{valero-lara_comparing_2023} conducted a comparative study of \mbox{LLaMA-2} and GPT-3 for generating numerical HPC kernels (AXPY, GEMV, and GEMM) across four programming languages. Their results showed that LLaMA-2 was generally less accurate, producing at least one correct implementation in 40\% of C++, 66\% of Fortran, 22\% of Julia, and 33\% of Python test cases, compared to GPT-3 which achieved 50\% in C++, 83\% in Fortran, 44\% in Julia, and 55\% in Python. However, when correct, LLaMA-2 frequently generated more optimized code than GPT-3. The main contribution in~\cite{valero-lara_comparing_2023} is its evaluation of optimization potential, although its relatively small and narrowly focused kernel set limits the generalizability of the findings to broader HPC applications.

Nichols et al. \cite{nichols_can_2024} evaluated generic LLMs such as GPT-3.5, GPT-4, StarCoder and CodeLlama in generating and scaling parallel code across diverse parallel programming models (OpenMP, MPI, and CUDA) and twelve problem domains. Their findings highlighted significant performance degradation in parallel tasks compared to serial code, with MPI posing particular challenges for current LLMs. From the execution model standpoint, OpenMP is the best-performing due to its similarity with serial code; on the other hand, MPI and MPI+OpenMP struggled due to higher complexity and different code structure. Efficiency was low for all models (<13\%), indicating suboptimal scaling despite correctness. While the benchmark provides a valuable first look at LLM limitations in HPC context, it does not explore targeted prompt-engineering or fine-tuning strategies that could mitigate poor MPI performance, leaving open questions about whether the observed deficits are intrinsic model limitations or dataset-related gaps (e.g., quality of the data respect to MPI code).

To address these limitations, Nichols et al. \cite{nichols_hpc-coder_2024} proposed HPC-Coder, a model fine-tuned specifically on HPC code. HPC-Coder demonstrated a 97\% success rate in generating functionally correct OpenMP pragmas, where correctness was defined as code that both compiles successfully and produces outputs consistent with the reference implementation under test cases, and achieved high accuracy in predicting the performance implications of code modifications (92\% on a coding competition dataset and 88\% on a proxy application dataset). It consistently outperformed generic LLMs in tasks such as loop parallelization and multi-language kernel generation---the improvement is particularly notable in MPI, where correctness increased by ~15\% over GPT-4. However, the model's evaluation was limited to benchmarks with relatively constrained code complexity (function-level tasks) and it remains unclear how HPC-Coder would generalize to highly heterogeneous, real-world HPC codebases without further domain adaptation.

Building on this foundation, Chaturvedi et al. \cite{chaturvedi_hpc-coder-v2_2024} introduced HPC-Coder-V2, which incorporated synthetic datasets (HPC-INSTRUCT) and task-specific fine-tuning. Here, task-specific fine-tuning refers to supervised instruction tuning on curated prompt-response pairs that target HPC actions (e.g., adding OpenMP pragmas, restructuring loops), rather than generic code chat. Evaluated on the ParEval benchmark \cite{nichols_can_2024}, HPC-Coder-V2 outperformed open-source models (e.g., StarCoder2, CodeLlama, Phind) roughly by 10-20 percentage points, and achieved near GPT-4 performance across all programming languages tested, when measuring the probability that the first code sample generated by an LLM passes all test cases for a given programming task (\texttt{pass@1} metric). This version emphasized the critical role of data quality and representation over sheer dataset volume. While synthetic data improved performance in low-resource settings, the reliance on generated rather than real-world HPC code raises concerns about potential overfitting to artificial patterns not representative of production HPC workloads.

Godoy et al. \cite{godoy_large_2024} assessed GPT-based tools such as GitHub Copilot and ChatGPT for kernel generation in C++, Fortran, Python and Julia. They reported high correctness for OpenMP and CUDA tasks but lower success for newer paradigms like HIP, emphasizing the influence of prompt specificity and the model's familiarity with programming languages on output quality. This highlights an important dependency on both prompt engineering skill and model pretraining exposure, but the lack of systematic control over prompt formulation reduces the reproducibility and comparability of results across models.

In a novel study, Valero-Lara et al. \cite{valero-lara_chatblas_2024} demonstrated the feasibility of creating an entire Basic Linear Algebra Subprograms (BLAS) library using LLMs without traditional hand-written kernels. By orchestrating a multi-agent system in which LLMs collaboratively performed roles like code writing, reviewing and optimization, they produced a fully functional and portable Level-1 BLAS library. Although these AI-generated libraries do not yet surpass manually optimized implementations in performance, they achieved between 70-90\% of the efficiency of expert-coded kernels, demonstrated portability across four programming languages (C++, Fortran, Python, Julia) and showed scalability with problem size close to linear, highlighting their potential for future HPC workflows. Nonetheless, the multi-agent orchestration process was highly curated by researchers, which suggests that such workflows are not yet fully autonomous or ready for unsupervised deployment.

When it comes to code translation, Ranasinghe et al.~\cite{ranasinghe_llm-assisted_2025} aim to assess ability of state-of-the-art LLMs to assist in translating legacy Fortran HPC codes into modern C++, across different models and platforms. The best performing GPT-4 produced syntactically and semantically valid translations in most attempts (90\% of the time providing semantically correct output).  Code generated both by GPT-4 and Claude runs near-native speeds, making LLM-assisted translation a feasible modernization strategy. Similarly to previous research, open-source models underperform unless heavily fine-tuned and guided. While the results are promising for legacy code migration, the study focuses on relatively clean, structured Fortran sources, leaving open the question of performance on messy, undocumented or domain-specific legacy codebases often encountered in practice.

\reviewSI{Pophale et al. \cite{Pophale2026ChatPORT} introduce ChatPORT, a fine-tuned LLM specialized for code porting, with a particular emphasis on OpenMP offloading translation from existing HPC kernels. ChatPORT improves correctness by an average of 43.2\% over its untuned based models, with an additional 6\% gain when feedback-derived data is incorporated into training. The paper shows that even modest fine-tuning can substantially improve translation fidelity in a narrow HPC setting, especially when the target programming model is underrepresented in general code corpora. At the same time, its scope is narrow: the evaluation centers on kernel-level portability rather than application-level integration, so the results say more about syntactic and semantic translation quality than about end-to-end performance on complex scientific software. Jin et al. \cite{Jin2025ChatPORTCudaSyCL} extend ChatPORT to CUDA-to-SYCL kernel translation, a practically important portability problem given the growing need to migrate away from vendor-specific GPU programming stacks. Their fine-tuning strategy yields executable SYCL code across all four trained code-model families and builds on a task where simple prompting is typically brittle because even small memory-model or launch-configuration mismatches break correctness. This is a meaningful contribution because it pushes LLM-based translation beyond directive rewriting into heterogeneous accelerator portability. However, the evaluation remains kernel-centric.}

\reviewSI{The Fortran2CPP work \cite{Chen2024Fortran2CPP} frames legacy Fortran modernization as a dialogue-driven translation problem, combining multi-turn prompting with dual agent interaction to improve syntactic validity and structural fidelity. The authors report up to a 3.31$\times$ improvement in CodeBLEU \cite{ren2020codebleumethodautomaticevaluation} and a substantial increase in compilation success for fine-tuned open-weight models, suggesting that translation quality can be improved when the LLM is guided through staged interaction rather than one-shot conversion. This is especially relevant for HPC because Fortran code often embeds domain conventions, numerical assumptions and build constraints that are difficult to recover from isolated prompts. However, the paper still relies heavily on translation-oriented metrics such as CodeBLEU and compilation rate, which are useful but do not fully resolve the question of whether translated codes preserve numerical behavior, performance characteristics and long-term maintainability in production environments.}

\reviewSI{Dubey and Dhruv \cite{DubeyDhruv2025NewCapability} address a less frequently studied but highly relevant scenario: using an LLM to implement genuinely new capability inside an existing scientific application, rather than merely translating or optimizing pre-existing code. Their contribution lies in proposing a methodology in which specification refinement, test-driven development and iterative prompting are used to guide the model toward implementing a new algorithmic feature in Flash-X \cite{Dubey_2022}. This paper exposes a different bottleneck than most code generation studies---namely, that the central challenge is often not syntax generation but making the problem specification precise enough for the model to act on. Its limitation is that the methodology remains labor-intensive and strongly dependent on human refinement and validation, which means it currently scales more as a reliable co-development workflow.}

\reviewSI{UniPar \cite{Bitan2025UniPar} proposes a unified framework for translating between serial, OpenMP and CUDA-style code paths, combining fine-tuning, inference-time hyperparameter tuning and compiler-guided repair in a single pipeline. The reported jump from roughly 46\% compilation / 15\% functional correctness for the default baseline to around 69\% compilation / 33\% correctness under the full UniPar workflow shows that translation quality depends not just on model choice but on the orchestration around the model. Still, the absolute correctness levels remain modest, which reinforces the idea that even sophisticated orchestration helps expose latent capability but does not yet eliminate the semantic fragility of LLM-generated HPC code.}


A notable distinction across the reviewed studies lies in the origin and specificity of the training data used. Most works, such as those by Nichols et al. \cite{nichols_can_2024}, Valero-Lara et al. \cite{valero-lara_comparing_2023,valero-lara_chatblas_2024}, Godoy et al. \cite{godoy_large_2024}, and Ranasinghe et al. \cite{ranasinghe_llm-assisted_2025}, evaluate pre-trained general-purpose or code-focused LLMs (e.g., GPT-4, StarCoder, LLaMA-2) whose training corpora draw heavily from broad, public code repositories with possible, undisclosed proprietary curation, without explicit inclusion of HPC-specific content. In contrast, HPC-Coder \cite{nichols_hpc-coder_2024} and HPC-Coder-V2 \cite{chaturvedi_hpc-coder-v2_2024} were fine-tuned on curated or synthetic parallel programming datasets, including OpenMP pragma labeling, HPC benchmarks and the HPC-INSTRUCT dataset. This targeted training appears to contribute directly to their superior performance, particularly in programming paradigms that are low-resource in the sense of being underrepresented in public datasets (notably SYCL and OpenCL), as well as more complex paradigms such as MPI, underscoring the importance of domain-specific data in advancing LLM-assisted HPC code generation.

Despite notable progress, several challenges persist in the field of LLM-driven parallel code generation:

\begin{itemize}
    \item Data limitations: Pretraining datasets often lack high-quality and diverse parallel code samples. While synthetic data generation mitigate this to an extent \cite{nichols_hpc-coder_2024,chaturvedi_hpc-coder-v2_2024}, it risks introducing artifacts may not capture real-world complexities.
    \item Generalization issue: LLMs struggle to adapt across varied execution models, especially for less mainstream paradigms such as HIP and for more complex programming models such as MPI, CUDA, etc. Godoy et al. \cite{godoy_large_2024} observed that GPT-based models achieved high correctness in OpenMP and CUDA but performed poorly for HIP kernels. Similarly, Nichols et al. \cite{nichols_can_2024} reported a substantial drop in performance for MPI and OpenMP+MPI compared to OpenMP-only tasks, citing the increased complexity and structural differences of distributed-memory programming. Beyond correctness, these shortcomings directly affect how generated code performs on final hardware architectures, where communication overheads, memory hierarchies, and workload scaling determine actual efficiency.
    \item Tuning strategies: There is limited understanding of how different instruction-tuning techniques impact generation quality for HPC-specific tasks. Chaturvedi et al. \cite{chaturvedi_hpc-coder-v2_2024} demonstrated that task-specific instruction tuning on the HPC-INSTRUCT dataset significantly improved performance. However, comparative results across different tuning approaches remain scarce, and Valero-Lara et al. \cite{valero-lara_chatblas_2024} noted that even multi-agent orchestration for BLAS library generation still required substantial manual guidance to ensure correctness and efficiency.
    \item \reviewSI{Translation fidelity and maintainability: Recent work highlights that generating syntactically correct code is insufficient for practical HPC adoption. LLM-generated translations \cite{Chen2024Fortran2CPP, Jin2025ChatPORTCudaSyCL} may compile and pass tests but still fail to preserve numerical stability, performance characteristics or long-term maintainability. Multi-turn and agent-based approaches improve structural fidelity, but they introduce additional complexity and reliance on human validation, limiting scalability.}
\end{itemize}

\reviewSI{Beyond general code translation, Fortran represents a particularly critical
frontier for LLM-assisted modernization in HPC. Large, long-lived application
codes in climate, astrophysics, and engineering remain predominantly written in
Fortran, yet the pool of expert Fortran developers is shrinking and there is
growing concern about long-term maintainability and skills gaps in HPC
centres~\cite{morgan_skills_2023}. Recent LLM-based efforts explicitly
target this space, proposing workflows for Fortran-to-C++ translation and
incremental refactoring that combine multi-turn dialogue, dual-agent
architectures, and extensive regression testing to preserve scientific
correctness~\cite{DhruvDubey2025CodeTranslation,Chen2024Fortran2CPP}. These
studies suggest that LLMs can substantially accelerate the mechanical aspects
of modernization (e.g., API translation, boilerplate restructuring), but also
highlight that human domain expertise and strong test harnesses remain
indispensable to avoid subtle numerical or performance regressions in legacy
Fortran codes.}

\subsection{Parallelization and Optimization}\label{subsec:parallel}
Parallelization and optimization are fundamental challenges in HPC, where transforming code for concurrent execution and achieving scalability across architectures are critical for efficiency. Research in this category explores the potential of LLMs to automate these tasks across diverse parallel programming paradigms, including OpenMP, CUDA, MPI, and various task-based models (e.g., OpenCL, SYCL, etc.) 

LLMs have demonstrated varying degrees of success in automating parallelization workflows and performance tuning. For instance, AutoParLLM, developed by Mahmud et al. \cite{mahmud_autoparllm_2025}, employs GNNs to enhance prompt engineering for context-aware parallel code generation in OpenMP. 
The proposed approach demonstrates notable gains in both code quality and execution performance, improving correctness by 11.7\% relative to baseline LLM outputs and attaining a 2.3× speedup, surpassing the 1.6× speedup achieved by standard LLMs. The gains are compelling, but the evaluation centers on OpenMP kernels with relatively contained scope; it remains unclear how the GNN-guided prompting scales to heterogeneous models (e.g., GPUs) or distributed memory (MPI), where data dependencies and synchronization are more complex.

In the domain of optimization, LLM-powered agent systems, such as the framework proposed by Wei et al.~\cite{wei_improving_2025}, leverage structured domain-specific languages (DSLs) and enriched system feedback loops to iteratively refine code. 
These DSLs---tailored for expressing parallel patterns, loop transformations, and low-level optimizations---provide a constrained and interpretable search space that reduces hallucinations and facilitates systematic performance exploration. Moreover, the integration of iterative feedback loops enables the system to incorporate compiler diagnostics, runtime profiling, and performance analysis data (e.g., execution time, cache misses, scalability curves) into subsequent refinement steps, moving beyond static prompt-response generation toward a more dynamic, optimization-driven interaction. 
These systems often surpass human expert-tuned implementations in both efficiency and quality. Similarly, domain-specific models like HPC-Coder \cite{nichols_hpc-coder_2024} and HPC-GPT \cite{ding_hpc-gpt_2023} (both fine-tuned on HPC source code) have excelled in tasks such as pragma labeling and predicting the performance impacts of code transformations. While the agent framework can surpass expert-tuned code on the chosen tasks, it relies on a carefully engineered DSL and iterative feedback loops that may be costly to integrate into HPC CI/CD pipelines. The approach's portability to large scientific applications is not yet demonstrated.

Cui et al. \cite{cui_large_2025} investigated whether LLMs understand and can apply performance optimization principles in C, C++ and Fortran code. The paper introduces OptiCode, a benchmark to evaluate LLMs' ability to generate, recognize and explain performance-optimized code across key categories such as loop fusion, unrolling, blocking, vectorization and memory reuse. Results show that GPT-4 consistently outperforms others in all optimization categories (with at least +15\% advantage in correct solutions). Open models (like StarCoder) fall significantly behind, especially in blocking and unrolling.

In~\cite{abdullah_p4omp_2025}, Abdullah et al. presented P4OMP, a retrieval-augmented generation (RAG) framework for transforming serial C/C++ code into OpenMP-parallelized code using LLMs. P4OMP integrates domain-specific tutorial knowledge into LLM prompts to improve both syntactic correctness and semantic fidelity during code transformation, without requiring fine-tuning. P4OMP succeeded in all parallelizable cases (100\% compilation success), compared to baseline GPT-3.5 which failed in 20 out of 108 cases (mainly due to missing or invalid variable scoping, syntax errors in OpenMP pragmas, and misuse of reduction clauses and iterators). Although the perfect compliance rate highlights the value of RAG in mitigating common semantic errors, it also raises the possibility of overfitting to the retrieved tutorial knowledge, since the benchmark scope was relatively narrow and closely aligned with the retrieval corpora. Thus, although RAG eliminates the need for fine-tuning and achieves strong compile success, the approach is tied to OpenMP tutorials and C/C++; it does not yet address GPU offload or distributed-memory models, and baseline failures (e.g., scoping, reductions) suggest persistent semantic gaps without retrieval.

\reviewSI{ChatMPI \cite{ValeroLara2026ChatMPI} extends the assistant-based paradigm specifically to MPI. By focusing on six representative HPC workloads and explicitly encouraging the model to reason about decomposition and communication patterns, the authors report higher trustworthiness than untuned baselines and show that the resulting codes can achieve up to a 4$\times$ performance boost over other AI generated variants. This work moves beyond syntax-level MPI generation toward communication-aware parallelization, which is precisely where many general-purpose models fail. However, the assistant is still evaluated on a relatively small set of structured kernels and its reliance on carefully curated task-specific training means it remains unclear how robust the approach would be for irregular communication patterns or larger MPI applications.}

\reviewSI{PEAK \cite{Tariq2025PEAK} reframes GPU kernel optimization as a sequence of natural language transformations, supported by validation and performance checks, rather than as monolithic end-to-end code generation. This is a strong conceptual advance because it turns optimization into transparent and modular process that can be steered by humans or agents while retaining portability across CUDA, HIP and even HLSL backends. The reported competitiveness with vendor libraries on matrix multiplication workloads suggest that the approach is not just explanatory but can produce genuinely useful low-level performance engineering outcomes. The main caveat is that PEAK's success currently rests on a curated transformation library and a narrowly defined workload family, so its transferability to more diverse GPU kernels and less structured optimization problems remains to be shown.}

\reviewSI{PRAGMA \cite{Lei2025PRAGMA} pushes optimization further by introducing a profiling-reasoned multi-agent systems in which low-level performance feedback, rather than correctness alone, guides kernel refinement. Evaluated on KernelBench \cite{ouyang2025kernelbenchllmswriteefficient}, the framework achieves a sizable average speedups over both naive baselines and untuned Torch implementations, showing the practical value of closing the loop between code generation and profiling data. This is one of the clearest demonstrations that LLM-based optimization can become more architecture-aware when profiling feedback is treated as first-class input. At the same time, the framework's complexity is also its main limitation: multi-agent coordination, profiling infrastructure and iterative repair all raise integration costs, which may make the approach harder to deploy outside the controlled performance-engineering settings.}

\reviewSI{Recent work has begun to extend LLM-driven optimization beyond performance metrics toward energy-aware computing. In this context, Dearing et al. \cite{Dearing2025LASSIEE} explore how LLMs can assist in transforming existing parallel programs to reduce energy consumption while preserving correctness and performance characteristics. The approach integrates LLM-based code transformation with feedback derived from execution metrics, enabling model to suggest refactoring such as loop restructuring, synchronization adjustments or memory access optimizations that impact both runtime and energy usage. The study also highlights a fundamental limitation: the effectiveness of such refactoring depends heavily on the availability and quality of runtime feedback, as well as on accurate modeling of hardware-specific energy behavior. As a result, while LLMs can propose plausible transformations, their ability to consistently produce energy-optimal code remains constrained without tight integration with profiling tools and architecture-aware analysis.}

Persistent obstacles in this domain include: 
\begin{itemize}
    \item LLMs occasionally generate syntactically correct but semantically flawed or inefficient code. Beyond typical issues such as mishandling shared memory, this also includes hallucinated functions, reliance on deprecated APIs or outdated versions, or incorrect parameter usage that produces runtime errors. P4OMP shows GPT-3.5 often compiles but fails due to invalid variable scoping, incorrect reduction use and pragma syntax errors; the RAG method fixed many of these issues \cite{abdullah_p4omp_2025}. OptiCode highlights that models frequently miss blocking/unrolling/vectorization opportunities---producing functionally correct yet performance suboptimal code \cite{cui_large_2025}. Domain-tuned models help, but their evaluations are still limited to controlled settings~\cite{nichols_hpc-coder_2024,ding_hpc-gpt_2023}.
    \item Large HPC source files frequently exceed token limits, constraining model usability. Methods in this category typically operate on short kernels or single-file snippets and avoid repository-level transformations, reflecting context-window constraints \cite{mahmud_autoparllm_2025,abdullah_p4omp_2025}. Agentic approaches mitigate this by chunking and DSL-structuring work, but evidence on long-context scientific codes is still lacking in these studies \cite{wei_improving_2025}.
    \item Most evaluations focus heavily on shared-memory paradigms (like OpenMP) \cite{mahmud_autoparllm_2025,abdullah_p4omp_2025}, with insufficient testing for distributed-memory approaches such as MPI or hybrid programming models. In distributed-memory contexts, additional problems arise from race conditions, deadlocks and non-deterministic behavior, which LLMs rarely address explicitly. HPC-Coder reports its strongest gains on OpenMP pragmas, with only partial improvements on MPI \cite{nichols_hpc-coder_2024}. HPC-GPT likewise emphasizes pragma/optimization reasoning more than distributed messaging \cite{ding_hpc-gpt_2023}. Together, these point to a skew toward shared-memory evaluations, with relatively sparse, rigorous testing on MPI or hybrid scenarios. It should be noted that even within shared-memory settings, emerging heterogeneous CPU designs complicate thread pinning and scheduling. Without explicit architectural context in prompts or retrieval, LLMs may select suboptimal strategies.
    \item \reviewSI{Recent approaches to performance and energy optimization
    demonstrate that LLMs are most effective when tightly integrated with compiler
    diagnostics, profiling data, and runtime or energy feedback
    \cite{Tariq2025PEAK,Lei2025PRAGMA,Dearing2025LASSIEE}. However, this creates a
    strong dependency on external tools and infrastructure, increasing system
    complexity. In the absence of such feedback, LLMs typically miss critical
    optimization opportunities and have no intrinsic understanding of
    energy–performance trade-offs, limiting their usefulness for meaningful
    optimization.}
\end{itemize}

\subsection{Framework and Architectures}\label{subsec:framework}
This category highlights innovations in the design and deployment of LLM frameworks tailored for HPC environments. These efforts aim to bridge the usability gap between domain scientists and complex HPC infrastructures by enabling programmable, conversational agents powered by LLMs. Emerging solutions emphasize modular architectures that incorporate fine-tuning strategies, RAG, real-time ingestion of user-specific contexts, privacy-preserving workflows and secure deployment pipelines. 

The chatHPC framework introduced by Yin et al. \cite{yin_chathpc_2024} exemplifies an end-to-end HPC-to-LLM system. It combines model pretraining and fine-tuning through self-improved self-instruct (SI2) hybrid RAG-based inference mechanisms and a containerized deployment setup with APIs and user interfaces for seamless interaction. Notable innovations include low-rank adaptation (LoRA) for efficient parameter updates, communication compression (ZeRO++) to optimize distributed training and advanced evaluation techniques such as hallucination detection, semantic coherence scoring and privacy auditing metrics. While chatHPC demonstrates a strong integration of LLM capabilities into HPC workflows, its reliance on synthetic benchmarks and lack of large-scale real-world deployment data limit the generalizability of its reported results. \reviewSI{A follow-on ChatHPC paper \cite{ValeroLara2025ChatHPC} broadens the earlier framework perspective by positioning LLMs not just as chat interfaces but as a family of small, specialized assistants embedded into an HPC software ecosystem. Its strongest contribution is architectural and operational: it demonstrates that trustworthy, task-specific assistants can be created with relatively small datasets and modest compute budgets for components such as programming-model translation, I/O libraries, math libraries and profiling tools. In that sense, it strengthens the framework narrative in this survey by showing that the field is moving from single-model prototypes toward ecosystems of targeted assistants. The trade-off, however, is that trustworthiness here is achieved through strong specialization and expert supervised data curation, so the resulting assistants may generalize poorly outside the narrow functions they were trained for.}

In parallel, Miyashita et al. \cite{miyashita_llm_2024} propose the HyCE-augmented RAG architecture, which extends conventional retrieval pipelines by embedding executable HPC command descriptions. This allows the system to produce contextually relevant and personalized responses. This adaptive architecture also incorporates automatic synthetic question-answer pair generation and iterative LLM-based evaluation loops to refine system accuracy and reliability over time. However, HyCE's effectiveness is highly dependent on the quality and coverage of the embedded command descriptions and its adaptability to HPC environments with non-standard tooling or workflows remains unclear.

Despite these promising developments, several challenges remain in designing robust frameworks and architectures for LLMs in HPC:

\begin{itemize}
    \item Integrating real-time, user-specific and resource manager systems data (e.g., active jobs, GPU availability) into LLM prompts is hindered by the heterogeneity of HPC environments and strict access controls---chatHPC \cite{yin_chathpc_2024}, for instance, requires environment-specific connectors, which may not generalize across all HPC systems.
    \item Ensuring privacy and safe command execution in multi-tenant HPC systems presents challenges in resource management and isolation when scaling LLM services---both frameworks \cite{yin_chathpc_2024,miyashita_llm_2024} acknowledge the need for sandboxing mechanisms but do not provide production-level implementations.
    \item The reliance on synthetic or self-generated evaluation datasets limits the ability of these frameworks to capture edge cases or generalize effectively across diverse tasks---HyCE's Q\&A generation method \cite{miyashita_llm_2024}, for example, may miss rare operational scenarios encountered in live HPC environments.
    \item Another consideration is the resource governance and cost awareness. Production deployments must respect allocation policies (queues, job credits, project budgets) and minimize inference cost. Frameworks should expose policy-aware prompts and agents bounded by quota, to keep LLM services predictable and affordable at datacenter scale.
    \item \reviewSI{Recent ecosystem-oriented approaches \cite{ValeroLara2025ChatHPC} improve reliability by focusing on narrow tasks, but this limits their ability to generalize across workflows or domains. Maintaining and scaling multiple specialized models also introduces operational overload.}
\end{itemize}

\subsection{Evaluation and Benchmarking}\label{subsec:evaluation}
This category encompasses research that systematically evaluates the capabilities of LLMs in generating, translating and optimizing HPC code. A recurring challenge identified across studies is that while general-purpose LLMs demonstrate strong proficiency with sequential code, they often struggle to handle the complexities of parallel programming paradigms effectively. 

The ParEval benchmark introduced by Nichols et al. \cite{nichols_can_2024} provides a comprehensive framework for assessing LLM performance on 420 parallel computing tasks across 12 computational problem types and seven programming models (including OpenMP, MPI, and CUDA). Key evaluation metrics presented in Table~\ref{tab:evaluation_metrics} were leveraged to quantify not only syntactic and semantic correctness, but also runtime performance and scalability. Although ParEval is broad and well-instrumented, most tasks are function-level and short context; results therefore understate challenges in repository-scale coordination (build systems, multi-file dependencies) and may overestimate generalization to production HPC codebases.

A continuation of ParEval, which focused on function-level tasks, is ParEval-Repo, a benchmark designed by Davis et al. \cite{davis_pareval-repo_2025} to evaluate LLMs on repository-level translation tasks from serial to parallel HPC code, covering diverse programming models and real-world repositories. Evaluation was conducted via unit tests, integration tests, runtime performances and human expert review. ParEval-Repo is a scalable, reproducible, and extensible benchmark with the goal of driving research on long-context and system-level code generation for HPC. While ParEval-Repo addresses long context realism, success remains uneven on larger repos, with recurring issues in build system generation, dependency wiring and interface consistency---indicating current LLMs still struggle with end-to-end software modernization at scale.

The chatHPC framework by Yin et al. \cite{yin_chathpc_2024} contributes a complete pipeline to fine-tune, serve and evaluate LLMs within HPC contexts. This system emphasizes the need for on-premise solutions to address security and performance constraints in HPC environments. The authors propose lexical and semantic evaluation schemes, coupled with hallucination detection mechanisms, to ensure reliability in generated outputs. The evaluation largely relies on synthetic datasets and framework-internal metrics; without extensive external studies or cross-site deployments, it is difficult to judge robustness under heterogeneous schedulers, modules and center-specific security policies.

Godoy et al. \cite{godoy_large_2024} evaluated GPT-based models for their ability to generate and auto-parallelize common scientific kernels using interactive prompting across multiple programming languages. The authors introduced a novel correctness metric, defined as the percentage of generated code snippets that not only compile successfully but also produce numerically accurate results when executed against predefined test cases. It enables comparative analysis across models and languages, revealing substantial disparities tied to both language maturity and model design.

In~\cite{nader_llm-2025}, Nader et al. evaluate the performance of the \mbox{DeepSeek-R1} LLM in generating code for five HPC benchmarks in four languages: C++, Fortran, Python, and Julia. The paper compares DeepSeek's performance against \mbox{GPT-4}, analyzing code correctness, compilation, execution and scalability. While DeepSeek demonstrates strong potential in reducing development time and generating valid code across domains, its generated HPC code suffers from scalability and optimization issues. The study highlights important model differences, but hardware/platform diversity (CPUs/GPUs, interconnects) is still limited.

\reviewSI{Dhruv and Dubey~\cite{DhruvDubey2025CodeTranslation} demonstrate one of the few large-scale evaluations of
LLM-assisted modernization, applying CodeScribe to a substantial Fortran
code base for LHC simulations. Their
workflow operates at application scale—spanning Fortran-to-C++ translation,
interoperability layers, and structural refactoring—while maintaining
correctness through supervised use and extensive testing.}

\reviewSI{PCEBench~\cite{Chen2025PCEBench} significantly enriches the benchmarking landscape by arguing that correctness alone is inadequate for evaluating LLMs in parallel code generation. The benchmark introduces a multi-dimensional metric suite---including compilation success, executability, data-race freedom, functional correctness and speedup over serial baselines---which better reflects the actual concerns of HPC developers than single-score pass metrics. Its limitation is that such a broad benchmark is inherently more complex to maintain and interpret, and the heavier evaluation stack may make reproducibility more difficult unless the toolchain and runtime environments are carefully versioned and shared.}

\reviewSI{The Mandelbrot study \cite{Diehl2025LLMHPCpp} offers a focused but useful application-level benchmark for evaluating LLM-generated C++, OpenMP and MPI+OpenMP code on a controlled workload. Its main value lies in moving evaluation beyond micro-kernels toward a complete, compile-run-debug pipeline where outputs can be checked both functionally and visually, while scalability is assessed across multiple parallelization strategies. This makes it a helpful bridge between function-level benchmarks such as ParEval \cite{nichols_can_2024} and the harder repository-level settings of ParEval-Repo \cite{davis_pareval-repo_2025}. At the same time, the benchmark's embarrassingly parallel structure also limits its generality.}

Despite substantial progress, the field continues to face critical challenges:

\begin{itemize}
    \item Effective evaluation requires realistic execution environments, proper integration with parallel libraries, and carefully selected input sizes---challenges not encountered in typical Python-based benchmarks. ParEval reports low efficiency and weak scaling even when code is functionally correct \cite{nichols_can_2024}. ParEval-Repo~\cite{davis_pareval-repo_2025} surfaces frequent build/integration failures and dependency issues at repo scale.
    \item A lack of architectural awareness in current LLMs limits their ability to optimize code for specific hardware configurations such as GPUs and multicore CPUs. Godoy et al.~\cite{godoy_large_2024} show correctness without competitive optimization for some languages/kernels. Nader et al.~\cite{nader_llm-2025} document scalability gaps on multi-core systems despite functional correctness.
    \item \reviewSI{Very few studies evaluate LLMs on large-scale, production scientific applications. Most experiments are confined to kernels, single solvers, benchmarks, or proxy apps, which under-represent integration complexity and long-running workflows. The Flash-X work by Dubey and Dhruv~\cite{DubeyDhruv2025NewCapability} comes closest to a real-application use case, but still relies on heavy human supervision, extensive testing, and carefully engineered scaffolding, underscoring that robust, largely autonomous LLM workflows for full production applications remain an open challenge.}
\end{itemize}

\subsection{Challenges and Landscape}\label{subsec:challenges}
This category encompasses research that provides critical contextualization of the state of integrating LLMs into HPC. Rather than proposing new technical frameworks, these works synthesize recent developments, outline persistent obstacles and suggest directions for future improvements. While emphasizing the potential of LLMs to enhance productivity, usability and performance in HPC workflows, they also illuminate the structural, technical and methodological barriers that must be overcome to achieve scalable and reliable adoption. 

Kommidi \cite{kommidi_bridging_2024} presents a broad empirical analysis of LLM applications in practical HPC settings, with case studies spanning domains such as climate science, genomics, aerospace, and healthcare. The study reports promising gains, including a reduction of up to 40\% in routine coding tasks, along with marked improvements in computational efficiency, including software compatibility issues, performance bottlenecks due to inference overhead and security concerns in enterprise environments. As potential solutions, the study recommends adopting modular architectures and developing domain-specific LLMs tailored to scientific domains.

In a more technical and conceptual analysis, Chen et al. \cite{chen_landscape_2024} provide a structured overview of current integration efforts. Their work identifies persistent gaps in areas such as dataset quality and diversity, program representation for parallel code and the lack of standardized evaluation frameworks. A notable contribution of the study is its in-depth critique of LLM performance in generating HPC code using parallel programming models such as OpenMP and MPI. The paper further explores the potential of multimodal systems and hybrid approaches that combine GNNs with LLMs to enhance context awareness and code understanding.

\section{Conclusions}\label{sec:conclusions}
LLMs have rapidly evolved from general-purpose natural language processors to specialized tools capable of supporting HPC workflows. This survey has systematically examined their application across five key categories: \reviewSI{(1) Code Generation, (2) Parallelization and Optimization, (3) Frameworks and Architectures, (4) Evaluation and Benchmarking, and (5) Challenges and Landscape. Tables~\ref{tab:paper-comparison}, ~\ref{tab:paper-comparison-2}, ~\ref{tab:paper-comparison-3} and ~\ref{tab:paper-comparison-4} summarize and compares contributions (column 4) and limitations (column 5) across these categories (column 1). The survey provides not only a snapshot of the state of the art but also a critical perspective on the maturity of this emerging field.}

\begin{table*}[htp]
\centering
\fontsize{9pt}{13pt}\selectfont
\caption{Comparison of included LLM-HPC research papers (1/4).}
\label{tab:paper-comparison}
\begin{tabular}{@{}p{0.6cm}p{1.9cm}p{4.0cm}p{4.8cm}p{4.8cm}@{}}
\toprule
\textbf{Cat.} & \textbf{Authors} & \textbf{Primary Focus} & \textbf{Main Contributions} & \textbf{Main Limitations} \\
\midrule
1, 4 & Nichols et al. (2024)~\cite{nichols_can_2024} & ParEval: function-level generation/scaling across models. & Broad cross-model baseline; exposes MPI difficulty and low efficiency $<$13\%. & Short-context tasks; limited repo-scale realism; no prompt/fine-tune ablations. \\
\arrayrulecolor[gray]{0.85}
\midrule
1, 2 & Nichols et al. (2024)~\cite{nichols_hpc-coder_2024} & HPC-Coder: pragma labeling, performance impact prediction. & 97\% correct OpenMP pragmas; +$\sim$15\% MPI correctness over GPT-4. & Evaluated on controlled benchmarks; generalization to large, messy repos unclear. \\
\midrule
1 & Chaturvedi et al. (2024)~\cite{chaturvedi_hpc-coder-v2_2024} & HPC-Coder-V2: synthetic HPC-INSTRUCT, task-specific tuning. & Strong gains over open-source model, near GPT-4 performance; data quality $>$ size. & Reliance on synthetic data; risk of overfitting to non-realistic patterns. \\
\midrule
1 & Valero-Lara et al. (2023)~\cite{valero-lara_comparing_2023} & Kernel generation (AXPY, GEMV, and GEMM). & Shows cases where LLaMA-2 is more optimized when correct. & Small kernel set; limited generalizability to broader apps. \\
\midrule
1, 2, 4 & Godoy et al. (2024)~\cite{godoy_large_2024} & Kernel generation \& auto-parallelization. & Multi-language, multi-model comparison; practical prompting. & Prompt variability; weaker results for HIP; limited reproducibility. \\
\midrule
1 & Valero-Lara et al. (2024)~\cite{valero-lara_chatblas_2024} & Multi-agent LLMs to build \mbox{Level-1} BLAS. & Feasible AI-generated library; decent portability. & Heavy human curation; performance below expert kernels. \\
\midrule
1 & Ranasinghe et al. (2025)~\cite{ranasinghe_llm-assisted_2025} & Fortran $\rightarrow$ C++ legacy translation. & High semantic validity (up to 90\%); near-native runtime. & Unclear performance on messy/underdocumented legacy code. \\
\midrule
2 & Mahmud et al. (2025)~\cite{mahmud_autoparllm_2025} & AutoParLLM: GNN-aided prompting for OpenMP. & +11.7\% correctness; 2.3$\times$ speedup vs 1.6$\times$ baseline. & OpenMP-centric; unknown transfer to MPI/GPU. \\
\midrule
2 & Wei et al. (2025)~\cite{wei_improving_2025} & Agent system with DSL + feedback for optimization. & Surpasses expert-tuned code on tasks; systematic feedback. & Engineering overhead; unclear portability to large multi-file apps. \\
\midrule
2 & Ding et al. (2023)~\cite{ding_hpc-gpt_2023} & HPC-GPT: domain-adapted for pragma/opt. reasoning. & Improves pragma inference and performance reasoning. & Sparse MPI/GPU evidence; limited cross-architecture validation. \\
\midrule
2, 3, 4 & Yin et al. (2024)~\cite{yin_chathpc_2024} & chatHPC: end-to-end HPC-LLM pipeline. & Integrated stack (LoRA, ZeRO++); on-premises emphasis. & Synthetic evaluation bias; limited real-world deployment evidence. \\
\arrayrulecolor{black}
\bottomrule
\end{tabular}
\end{table*}

\begin{table*}[htp]
\centering
\fontsize{9pt}{13pt}\selectfont
\caption{Comparison of included LLM-HPC research papers (2/4)}
\label{tab:paper-comparison-2}
\begin{tabular}{@{}p{0.6cm}p{1.9cm}p{4.0cm}p{4.8cm}p{4.8cm}@{}}
\toprule
\textbf{Cat.} & \textbf{Authors} & \textbf{Primary Focus} & \textbf{Main Contributions} & \textbf{Main Limitations} \\
\midrule
3 & Miyashita et al. (2024)~\cite{miyashita_llm_2024} & HyCE-augmented RAG with command descriptions. & Personalized, context-rich outputs; adaptive retrieval. & Coverage depends on embedded command corpora. \\
\arrayrulecolor[gray]{0.85}
\midrule
2 & Cui et al. (2025)~\cite{cui_large_2025} & OptiCode: optimization motifs benchmark. & Clear gap analysis; GPT-4 $>$ open models by 15\%. & Micro-pattern focus; end-to-end app impact not shown. \\
\midrule
2 & Abdullah et al. (2025)~\cite{abdullah_p4omp_2025} & P4OMP: RAG for serial $\rightarrow$ OpenMP. & 100\% compiles on parallelizable cases; fixes scoping/reduction errors. & Tied to OpenMP/C/C++; no GPU/MPI; retrieval coverage limits. \\
\midrule
4 & Davis et al. (2025)~\cite{davis_pareval-repo_2025} & ParEval-Repo: repo-level translation/modernization. & Long-context, system-level realism; reproducible. & Frequent build/dependency failures; interface consistency issues. \\
\midrule
4 & Nader et al. (2025)~\cite{nader_llm-2025} & DeepSeek-R1 vs GPT-4 on HPC benchmarks. & Reduces development time; many valid solutions. & Scalability/optimization deficits vs GPT-4; limited hardware diversity. \\
\midrule
5 & Kommidi (2024)~\cite{kommidi_bridging_2024} & Empirical cross-domain study (climate, genomics, etc.) & Practical orientation; proposes concrete integration directions. & Lacks deep technical analysis; high-level case studies. \\
\midrule
5 & Chen et al. (2024)~\cite{chen_landscape_2024} & Landscape review: gaps \& directions. & Technical \& conceptual review; highlights standardization needs. & Lacks empirical validation. \\
\midrule
2 & Kadosh et al. (2023)~\cite{kadosh_scope_2023} & Code understanding and completion. & Significant perplexity drop; outperforms BPE tokenizer. & Limited, early stage work; lacks full task benchmarking. \\
\arrayrulecolor{black}
\bottomrule
\end{tabular}
\end{table*}

\reviewSI{\begin{table*}[htp]
\centering
\fontsize{9pt}{13pt}\selectfont
\caption{Comparison of included LLM-HPC research papers (3/4).}
\label{tab:paper-comparison-3}
\begin{tabular}{@{}p{0.6cm}p{1.9cm}p{4.0cm}p{4.8cm}p{4.8cm}@{}}
\toprule
\textbf{Cat.} & \textbf{Authors} & \textbf{Primary Focus} & \textbf{Main Contributions} & \textbf{Main Limitations} \\
\midrule
1, 5 & Dubey and Dhruv (2025)~\cite{DubeyDhruv2025NewCapability} & LLM-assisted addition of new capability to Flash-X. & Shows structured methodology for implementing new algorithmic functionality with specification refinement. & Heavy human supervision; evidence is process-oriented rather than large-scale comparative evaluation. \\
\arrayrulecolor[gray]{0.85}
\midrule
1, 3 & Pophale et al. (2025)~\cite{Pophale2026ChatPORT} & Fine-tuned LLM for OpenMP offload translation. & +43.2\% correctness over base models; feedback-derived data further improves translation quality. & Kernel-level scope; translation correctness not tied to end-to-end application performance. \\
\midrule
1, 3 & Jin et al. (2025)~\cite{Jin2025ChatPORTCudaSyCL} & CUDA-to-SYCL kernel translation via parameter-efficient tuning. & Produces executable translated kernels across several code-model families; extends portability beyond directive rewriting. & Small/narrow evaluation set; executable SYCL does not necessarily imply optimized SYCL. \\
\midrule
1 & Chen et al. (2024)~\cite{Chen2024Fortran2CPP} & Multi-turn, dual agent Fortran-to-C++ translation. & Up to 3.31$\times$ CodeBLEU improvement and major gains in compilation success. & Translation metrics may overstate semantic or performance fidelity on real legacy HPC codes. \\
\midrule
1, 5 & Dhruv and Dubey (2025)~\cite{DhruvDubey2025CodeTranslation} & Assisted scientific software translation and interoperability. & Moves beyond kernels to supervised real-code workflows, including translation, APIs and code inspection. & Strong dependence on human oversight; demonstrates assisted modernization more than autonomous conversion. \\
\midrule
1, 3 & Bitan et al. (2025)~\cite{Bitan2025UniPar} & Unified framework for serial/OpenMP/CUDA translation. & Combines fine-tuning, hyperparameter tuning and compiler-guided repair; roughly doubles baseline translation quality. & Absolute functional correctness rates remain modest; long-kernel and broader API support still limited. \\
\midrule
1, 2, 3 & Valero-Lara et al. (2026)~\cite{ValeroLara2026ChatMPI} & MPI assistant for sequential C-to-MPI transformation. & Improves trustworthiness and can yield up to 4$\times$ performance gains via better decomposition/communication. & Evaluated on a small workload family; robustness for irregular or multi-file MPI codes remains unclear. \\
\arrayrulecolor{black}
\bottomrule
\end{tabular}
\end{table*}}

\reviewSI{\begin{table*}[htp]
\centering
\fontsize{9pt}{13pt}\selectfont
\caption{Comparison of included LLM-HPC research papers (4/4).}
\label{tab:paper-comparison-4}
\begin{tabular}{@{}p{0.6cm}p{1.9cm}p{4.0cm}p{4.8cm}p{4.8cm}@{}}
\toprule
\textbf{Cat.} & \textbf{Authors} & \textbf{Primary Focus} & \textbf{Main Contributions} & \textbf{Main Limitations} \\
\midrule
3 & Valero-Lara et al. (2025)~\cite{ValeroLara2025ChatHPC} & Ecosystem of specialized trustworthy HPC assistants. & Demonstrates small data, modest cost creation of task-specific assistants for multiple HPC components. & Strong specialization may limit transferability; trustworthiness depends on curated expert supervision. \\
\arrayrulecolor[gray]{0.85}
\midrule
2, 3 & Tariq et al. (2025)~\cite{Tariq2025PEAK} & Natural language GPU-kernel optimization assistant. & Modular transformation pipeline competitive with vendor libraries on matrix multiplication; portable across CUDA/HIP/HLSL. & Relies on curated transformations and narrow workload family; broader generality not yet established. \\
\midrule
2, 3 & Lei et al. (2025)~\cite{Lei2025PRAGMA} & Profiling-reasoned multi-agent kernel optimization. & Introduces closed-loop optimization with profiler feedback; sizable speedups on KernelBench. & Multi-agent and profiling stack increase deployment complexity; evaluation still controlled. \\
\midrule
4 & Chen et al. (2025)~\cite{Chen2025PCEBench} & Multi-dimensional benchmark for parallel code generation. & Extends evaluation beyond pass metrics to compilation, execution, race freedom, correctness and speedup. & Benchmark complexity raises maintenance and reproducibility burden; harder to compare than single-score suites. \\
\midrule
1, 4 & Diehl et al. (2025)~\cite{Diehl2025LLMHPCpp} & Mandelbrot-based evaluation of generated C++, OpenMP, MPI+OpenMP. & Application-level compile-run-debug benchmark bridging kernel and repository evaluations. & Workload is embarrassingly parallel, so generalization to communication-heavy HPC codes is limited. \\
\midrule
2 & Dearing et al. (2025)~\cite{Dearing2025LASSIEE} & Energy-aware LLM-driven refactoring of parallel codes. & Introduces energy-aware transformations guided by runtime feedback; aligns optimization with sustainability goals. & Strong dependence on profiling accuracy and hardware-specific models; limited generalization without closed-loop integration. \\
\arrayrulecolor{black}
\bottomrule
\end{tabular}
\end{table*}}

Recent research reveals that while general-purpose LLMs can provide reasonable support for serial programming tasks, their effectiveness declines sharply when applied to more complex parallel and distributed programming models. For example, Nichols et al. \cite{nichols_can_2024} show that models such as GPT-4 achieve relatively strong performance in OpenMP, which shares structural similarities with sequential code, yet their outputs degrade significantly in MPI-based tasks, where message passing and synchronization present far greater semantic challenges. These limitations underscore that syntactic correctness does not necessarily translate into functional code in HPC contexts.

Fine-tuned and domain-specific LLMs, such as HPC-Coder \cite{nichols_hpc-coder_2024}, HPC-GPT \cite{ding_hpc-gpt_2023} and chatHPC \cite{yin_chathpc_2024} demonstrate improved performance, particularly when augmented with targeted instruction tuning, domain-specific datasets and RAG mechanisms. Their success highlights the central role of data quality and specialization: When training corpora include high-quality parallel code, models not only generate more syntactically valid outputs but also achieve closer alignment with HPC performance objectives such as scalability and resource efficiency. However, these systems remain limited in scope, with evaluations often restricted to controlled benchmarks or relatively small code kernels. The question of whether they can generalize to the messy, heterogeneous workloads typical of production HPC centers remains unresolved.

Benchmarking efforts like ParEval \cite{nichols_can_2024} and ParEval-Repo \cite{davis_pareval-repo_2025} are another significant step forward. They provide standardized tools for assessing LLM capabilities across diverse programming models and problem domains, moving the field toward greater reproducibility and comparability. Metrics such as \texttt{pass@k}, runtime efficiency and scalability enable more nuanced evaluation than generic NLP scores like BLEU, which measures n-gram overlap with a reference; while useful for text, it correlates weakly with program semantics and runtime behavior. Yet, benchmarking itself is not without problems. A recurring concern, noted by Nichols et al. \cite{nichols_can_2024} and Davis et al. \cite{davis_pareval-repo_2025}, is that some LLMs may have been trained on datasets that overlap with benchmark tasks, artificially inflating performance. This risk of benchmark contamination highlights the need for continual renewal of datasets and the development of benchmarks that capture real-world challenges such as multi-file dependencies, build-system complexity and debugging of non-deterministic errors. This phenomenon has led some scholars \cite{eriksson_trust-ai_2025} to argue that the current benchmarking ecosystem has entered a state of saturation, where incremental improvements may reflect dataset familiarity rather than architectural innovation. Ultimately, the sustainability of benchmarking will depend on community-driven efforts to maintain transparency, dataset versioning and periodic renewal of the tasks used to measure LLM performance.

Beyond these technical insights, the broader narrative emerging from the literature is one of dual potential and fragility. On one side, LLMs promise to democratize HPC by lowering barriers to entry, assisting non-expert programmers and accelerating prototyping. On the other, they remain brittle when faced with the full demands of HPC, where correctness, scalability and performance cannot be compromised. The road from promising prototypes to trustworthy production tools will require advances in multiple areas: Richer domain datasets, hybrid AI–HPC integration frameworks, rigorous and uncontaminated evaluation practices and methodologies for human-in-the-loop validation. \reviewSI{Newer work also shows that the field is no longer dominated solely by kernel generation studies: there is an increased emphasis on specialized translation assistants, profiling-guided optimization loops and ecosystem-level deployments of small, trustworthy LLM agents tailored to particular HPC components.}

Together, these findings demonstrate that LLMs are unlikely to replace HPC experts in the foreseeable future but are poised to become powerful assistants in the HPC software development pipeline. Their successful deployment will depend not only on technical innovations but also on the creation of an ecosystem - benchmarks, datasets, frameworks and governance structures - that ensures transparency, accountability and sustainability. Far from a finished story, LLM-HPC integration should be seen as an evolving collaboration between AI and scientific computing, where each step forward uncovers new challenges but also new opportunities for reshaping the future of high-performance software development.

\subsection{Answer to Research Questions}
\paragraph{RQ1: In which ways have LLMs been utilized in HPC domain so far?}

LLMs have been deployed along the entire HPC software lifecycle. At the \textbf{code level}, they generate kernels and insert pragmas (e.g., OpenMP) from natural language or serial code, translate legacy Fortran to C++ while preserving semantics and performance and propose micro-optimizations (e.g., loop transforms) \cite{nichols_can_2024,nichols_hpc-coder_2024,ranasinghe_llm-assisted_2025}. At the \textbf{parallelization level}, they suggest execution models (OpenMP vs. CUDA vs. MPI), label parallel regions, and attempt to restructure loops and memory access patterns \cite{mahmud_autoparllm_2025,nichols_hpc-coder_2024}. At the \textbf{system/workflow level}, LLMs appear as conversational agents embedded in frameworks that understand schedulers, modules and site-specific tooling (e.g., chatHPC, HyCE), enabling context-aware answers about jobs, queues and environment setup \cite{yin_chathpc_2024,miyashita_llm_2024}. There are also \textbf{multi-agent pipelines} that coordinate roles such as authoring, reviewing and refactoring (e.g., ChatBLAS for \mbox{Level-1} BLAS generation), and \textbf{benchmark-centric uses} where LLMs are evaluated across parallel models and repositories (ParEval / ParEval-Repo) \cite{valero-lara_chatblas_2024, nichols_can_2024, davis_pareval-repo_2025}. In short, use spans from “assistive pair-programmer” to “pipeline component” in integrated HPC environments, though most production-ready benefits today cluster around guided generation and translation rather than autonomous end-to-end development. \reviewSI{In addition to code generation and optimization, newer work shows growing use of LLMs for code translation and software modernization (\cite{Chen2024Fortran2CPP, Jin2025ChatPORTCudaSyCL} and for specialized assistant ecosystems that support profiling, portability, I/O and performance engineering.}

\reviewSI{In practical terms, one of the most impactful near-term application areas for
LLMs in HPC is likely to be the maintenance and modernization of legacy
Fortran-based applications, where declining language expertise and large,
mission-critical code bases make assistive translation and refactoring tools
particularly valuable~\cite{DhruvDubey2025CodeTranslation,Chen2024Fortran2CPP}.}

\paragraph{RQ2: Based on current research, what are the primary techniques and modes used to leverage LLMs for HPC?}
\reviewSI{Five modes stand out:}
\begin{enumerate}
    \item \textbf{Domain fine-tuning} on curated HPC corpora (e.g., HPC-Coder, HPC-Coder-V2), often with instruction-style data (HPC-INSTRUCT) that materially improves low-resource domains like SYCL/OpenCL and raises MPI correctness \cite{nichols_hpc-coder_2024,chaturvedi_hpc-coder-v2_2024}.
    \item \textbf{Instruction tuning} with task-specific prompts (pragma labeling, kernel refactoring, speedup prediction) to align model behavior with HPC semantics~\cite{nichols_hpc-coder_2024,chaturvedi_hpc-coder-v2_2024}.
    \item \textbf{Retrieval-augmented generation (RAG)} that injects authoritative context (tutorials, command manuals, site docs), reducing hallucinations and fixing common pragma/scoping mistakes without fine-tuning (e.g., P4OMP; HyCE’s executable command descriptions)~\cite{abdullah_p4omp_2025,miyashita_llm_2024}.
    \item \textbf{Agentic/Hybrid approaches}: Multi-agent role splitting (ChatBLAS), graph-based guidance (AutoParLLM’s GNN-augmented prompts) and DSL-constrained search with iterative system feedback (agentic optimizers)~\cite{valero-lara_chatblas_2024,mahmud_autoparllm_2025,wei_improving_2025}. Strengths vary: Fine-tuning and instruction tuning deliver the largest accuracy jumps but need high-quality data; RAG is lightweight and portable; agentic/DSL methods trade engineering overhead for robustness and auditability.
    \item  \reviewSI{\textbf{Compiler-guided repair, multi-turn dialogue and profiling-guided agent loops} are emerging as important techniques, suggesting that successful HPC-LLM workflows increasingly depend on structured orchestration around the model rather than on raw prompting along. \cite{Chen2024Fortran2CPP, DhruvDubey2025CodeTranslation, Bitan2025UniPar, Lei2025PRAGMA}}
\end{enumerate}

\paragraph{RQ3: How effective are LLMs in assisting with parallel programming and HPC-related tasks?}

Effectiveness is paradigm-dependent and data-dependent. General-purpose models show reasonable capability on OpenMP tasks (structurally close to serial code) but struggle on MPI and hybrid models, where synchronization, data distribution and nondeterminism amplify semantic difficulty; ParEval reports low efficiency and weak scaling even when code is ``correct'' \cite{nichols_can_2024}. Domain-tuned models lift accuracy substantially---HPC-Coder achieves 97\% correctness for OpenMP pragma labeling and improves MPI correctness over GPT-4 by 15\% \cite{nichols_hpc-coder_2024}, and HPC-Coder-V2 narrows the gap in low-resource models via higher-quality instructions \cite{chaturvedi_hpc-coder-v2_2024}. RAG can close common semantic gaps (P4OMP fixed scoping/reduction issues and achieved 100\% compilation on parallelizable cases) \cite{abdullah_p4omp_2025}. For translation, GPT-4/Claude produced semantically correct Fortran to C++ mappings with near-native runtime in many cases \cite{ranasinghe_llm-assisted_2025}. Yet, optimization depth remains inconsistent: OptiCode shows open models miss blocking/unrolling/vectorization opportunities; even strong models require careful prompting and validation \cite{cui_large_2025}. 
\reviewSI{Furthermore, evidence of effectiveness at full-application scale is still scarce, with only case studies such as CodeScribe on a large Fortran LHC codebase, going beyond kernels or proxy apps~\cite{DhruvDubey2025CodeTranslation}.}

In summary, LLMs are effective accelerators for guided parallelization and modernization, but still need human-in-the-loop and/or performance feedback to meet HPC’s correctness \& scalability bar.

\paragraph{RQ4: What evaluation metrics and benchmarks are used to assess LLM-generated code for HPC?}

Beyond generic software metrics, HPC-aware works track compilation/\texttt{pass@k}, runtime speedup, parallel efficiency and scaling behavior across models (OpenMP/MPI/CUDA), languages and problem types \cite{nichols_can_2024}. ParEval-Repo ups the realism with repository-level tests: Unit + integration tests, build system correctness, runtime performance and expert review to capture long-context/repo-wide coordination \cite{davis_pareval-repo_2025}. OptiCode probes optimization literacy across loop fusion, unrolling, blocking, vectorization, and reuse in C/C++/Fortran \cite{cui_large_2025}. Frameworks like chatHPC add lexical/semantic metrics and hallucination checks \cite{yin_chathpc_2024}. Two cross-cutting issues persist: (i) hardware realism (results depend on CPU/GPU configs, interconnects, input sizes), and (ii) benchmark contamination---models possibly trained on tasks similar to evaluation, risking inflated scores and obsolescence; both ParEval and ParEval-Repo flag this risk and move toward harder, less ``seen'' tasks \cite{nichols_can_2024,davis_pareval-repo_2025}. \reviewSI{Recent work such as PCEBench \cite{Chen2025PCEBench} further broadens this space by introducing multi-dimensional metrics that distinguish compilation, executability, race-freedom, functional correctness and performance improvement, thereby addressing limitations of single-score correctness evaluation.}

\paragraph{RQ5: How do LLM-generated HPC solutions compare to human-written code in terms of correctness, efficiency, and scalability?}

On correctness, LLMs can approach or match humans for narrow, well-scoped tasks (e.g., pragma labeling, idiomatic OpenMP kernels, BLAS-level routines) and for guided translation where structure is preserved \cite{nichols_hpc-coder_2024,valero-lara_chatblas_2024,ranasinghe_llm-assisted_2025}. On efficiency/scalability, the gap widens: ParEval reports efficiency <13\% across models for parallel scaling; MPI remains the weak spot due to communication and synchronization complexity \cite{nichols_can_2024}. Agentic/DSL systems can rival expert baselines on chosen tasks \cite{wei_improving_2025}, but portability to large, heterogeneous scientific applications is unproven. In practice, LLMs reduce development time and help reach a correct baseline faster; human performance engineers still drive architecture-aware optimization (NUMA placement, cache tiling, GPU memory movement) and end-to-end scaling.

\paragraph{RQ6: What are the key challenges and limitations of using LLMs for HPC?}
They are enumerated below:
\begin{enumerate}
    \item Data scarcity/mismatch: Parallel and performance-critical code is underrepresented in pretraining; synthetic data helps but can introduce artifacts \cite{nichols_hpc-coder_2024,chaturvedi_hpc-coder-v2_2024}.
    \item Generalization gaps: Weak transfer to MPI/HIP/hybrid models; higher error rates on distributed semantics and less common APIs \cite{nichols_can_2024,godoy_large_2024}.
    \item Context limits: Long, multi-file scientific repos exceed token windows; many studies limit to short kernels or file-level tasks \cite{mahmud_autoparllm_2025,abdullah_p4omp_2025,davis_pareval-repo_2025}.
    \item Semantic fragility: Hallucinated calls, deprecated APIs, wrong params and subtle race/ordering bugs; RAG reduces but doesn’t eliminate these \cite{abdullah_p4omp_2025}.
    \item Hardware unawareness: Limited sensitivity to cache/NUMA/interconnect/topology; poor optimization choices for specific architectures \cite{godoy_large_2024,nader_llm-2025}.
    \item Evaluation fragmentation and possible benchmark contamination; limited real-world deployments \cite{nichols_can_2024,davis_pareval-repo_2025,yin_chathpc_2024}.
    \item Operational considerations: Privacy/security for on-premises use, sandboxed execution of generated code and integration with schedulers and modules \cite{yin_chathpc_2024,miyashita_llm_2024}.
\end{enumerate}

\reviewSI{Although the current ``LLM hype'' is unlikely by itself to make HPC researchers
substantially more inclined to release full application code or datasets---given
legal, intellectual-property, and sensitivity constraints---it does not preclude
meaningful use of LLMs in HPC. In practice, many models are shared as open or
restricted weights without disclosing the underlying training data, and this
pattern is likely to continue. For HPC, this points toward a future where impact
comes less from universal data sharing and more from (i) domain- or
facility-specific models that are fine-tuned on local code bases and offered as
internal services, and (ii) open-weights, HPC-oriented models that can be
adapted to particular architectures or applications without access to the
original pretraining corpora, complemented by targeted open-science efforts
around benchmarks, proxy applications, and curated code slices rather than
entire production codes.}

\paragraph{RQ7: What research gaps exist in the application of LLMs to HPC?}

First, \textbf{high-fidelity datasets}: Open, diverse HPC corpora with execution traces, scaling curves and energy/memory profiles across CPU/GPU and shared/distributed memory are still rare. Second, \textbf{instruction-tuning science} for HPC: Ablations on prompt style, curriculum construction (from serial to OpenMP to MPI) and robustness to messy legacy repos remain underexplored \cite{chaturvedi_hpc-coder-v2_2024}. Third, \textbf{long-context, repo-scale generation} and \textbf{build-system reasoning} (CMake/Spack/modules) is a major gap exposed by ParEval-Repo \cite{davis_pareval-repo_2025}. Fourth, \textbf{architectural awareness}: Integrating performance analysis tools, profilers and predictive models (for OpenMP/MPI behavior) to inform LLM decisions is in its early stages but promising \cite{mahmud_autoparllm_2025,wei_improving_2025}. Fifth, \textbf{benchmarking standards} that minimize contamination and capture real-world complexity (multi-file, multi-node, nondeterminism, error handling) are needed to compare methods fairly \cite{nichols_can_2024,davis_pareval-repo_2025}. Finally, \textbf{governance/operations}---traceability of generated code, human-in-the-loop policies and safe deployment patterns need community consensus before widespread production adoption \cite{yin_chathpc_2024,chen_landscape_2024}.

\subsection{Future Research Directions}
Based on the current landscape and observed limitations, several directions emerge for future work:

\begin{itemize}
    \item Domain-specific training data for parallel programming: One of the most common limitations is the lack of high-quality, representative training data for complex parallel programming paradigms, particularly for MPI and hybrid GPU/CPU systems. Future efforts should prioritize compiling domain-specific datasets drawn from real-world scientific applications---ideally sourced from open-source repositories or industry contributions. Enriching these datasets with execution traces, performance metrics and metadata could enhance fine-tuning processes and enable the development of more generalizable and capable HPC-focused models. In addition, creating benchmarks suites across diverse scientific domains and including performance data from different architectures (CPU, GPU, shared/distributed memory) and workload granularities would provide both training material and robust evaluation standards.
    \item Hybrid architectures for enhanced code understanding: As evident in recent work (e.g., AutoParLLM~\cite{mahmud_autoparllm_2025}), LLMs often lack deep semantic understanding of control structures and data dependencies, which is critical for effective parallelization. Integrating symbolic tools or GNNs with LLMs may enable more accurate modeling of program logic, enhancing their ability to perform transformations such as loop restructuring, memory optimization and dependency management.
    \item HPC-specific evaluation benchmarks: Many existing benchmarks rely on general NLP or software engineering metrics (e.g., BLEU, \texttt{pass@k}), which fail to capture the operational priorities of HPC. There is a need for evaluation frameworks that reflect real-world scientific computing goals---such as runtime efficiency, scalability, memory footprint, energy consumption, debugging success rates, and correctness under parallel execution. Establishing such metrics will allow for more rigorous and application-relevant comparisons across different LLM-based tools.
    \item Interactive coding environments: A recurring theme in the literature is the fragility and opacity of LLM-generated code. Future research should therefore explore the development of interactive development environments (IDE), or plugins for them, specific to HPC, where developers can inspect, correct, and refine LLM outputs in real time. Systems that incorporate compiler and execution feedback (runtime, memory), allow iterative refinement, and retain human control are expected to increase trust, reduce debugging time, and promote adoption within the HPC community. While general-purpose assistants such as GitHub Copilot integrated into VS Code have demonstrated the utility of LLMs in everyday software engineering, HPC programming introduces unique requirements (e.g., parallelism, performance portability, hardware-specific optimization) that justify the design of domain-tailored interactive tools.
    \item Repository-grounded assistants: Building RAG pipelines over local code, papers and API manuals to constrain generations to approved interfaces and site policies, enabling center-specific copilots that are safer than general chat.
    \item Closed-loop performance optimization: Integrate tracing/profiling tools (e.g., mpiP, HPCToolkit, Nsight) to feed runtime data back to LLM agents, enabling iterative refactoring guided by measured bottlenecks.
    \item Energy-aware and sustainable HPC workflows: Given the dual energy demands of LLM inference and HPC systems, sustainability should be a central consideration. LLMs could be integrated into energy-aware tooling that recommends energy-efficient algorithm variants, efficient resource allocation strategies or scheduling adjustments based on historical energy usage profiles, as well as to jointly optimize energy and monetary cost. Embedding LLMs into tools that promote energy efficiency can contribute meaningfully to the ecological footprint of scientific computing. \reviewSI{Recent work \cite{Dearing2025LASSIEE} has begun to explore this direction more concretely, demonstrating that LLMs can assist in energy-aware refactoring of parallel scientific codes by leveraging execution feedback to guide transformations that reduce energy consumption while preserving correctness and performance, although such approaches still rely heavily on accurate profiling and hardware-aware modeling.}
    \item Collaborative infrastructure and community standards: Future advances will benefit from cross-disciplinary collaboration and shared infrastructure. Participating in initiatives such as the Trillion Parameter Consortium (TPC)~\cite{noauthor_trillion_nodate} can help standardize best practices for model development, promote dataset sharing and align AI and HPC communities around common challenges like reproducibility, deployment safety and benchmarking rigor. Collective efforts are essential to ensure scalable, safe and scientifically grounded adoption of LLMs in HPC workflows.
    \item Establishing a neutral benchmarking consortium for LLM-HPC (similar to LMArena\footnote{\url{https://lmarena.ai}}) with rotating, held-out tasks and reproducible frameworks (evaluation suites) to prevent contamination and ensure comparability.
    \item \reviewSI{New systems such as PRAGMA \cite{Lei2025PRAGMA} and PEAK \cite{Tariq2025PEAK} suggest that the future progress may come from connecting LLMs to profilers, compilers and runtime monitors in a closed optimization loop. Rather than generating code once, such systems can iteratively reason over execution traces, hardware counters and validation results to refine correctness and performance.}
    \item \reviewSI{The rise of ChatPORT \cite{Pophale2026ChatPORT}, Fortran2CPP \cite{Chen2024Fortran2CPP} and UniPar \cite{Bitan2025UniPar} indicates a need for benchmarks that evaluate not only compilability and correctness, but also portability fidelity, API adaptation quality and preservation of performance-relevant semantics across source and target programming models.}
\end{itemize}

\subsection{Ethical, Social, and Environmental Impact}
The integration of LLMs into HPC workflows raises a range of ethical, social and environmental considerations that warrant careful evaluation.

Several ethical risks emerge from the application of LLMs in HPC contexts \cite{chen_landscape_2024}, \cite{guo_bias_2024}, \cite{jiao_navigating_2025}:

\begin{itemize}
    \item Bias propagation: LLMs may inherit and reproduce biases present in their training data, which could manifest in generated code or recommendations, particularly in sensitive scientific or policy-driven applications.
    \item Accountability: As LLM-generated code becomes more complex, tracing the origin of logic errors and ensuring transparent debugging becomes increasingly difficult. This opacity challenges conventional methods of accountability in scientific computing. Addressing this requires explicit methodologies for traceability, such as logging generated code with version metadata, maintaining links between prompts and outputs and adopting ``human-in-the-loop'' oversight mechanisms that ensure experts review and validate critical outputs.
    \item Proprietary dependence: The use of closed-source models (e.g., GPT-4) raises concerns around transparency, reproducibility and vendor lock-in. Relying on proprietary systems may limit long-term sustainability and control over critical scientific infrastructure. In addition, since benchmarking, reproducibility and performance validation are essential in HPC, open source approaches become particularly important; not only to mitigate vendor dependence but also to support accountability and community-driven verification.
\end{itemize}

The deployments of LLMs in HPC also introduces notable social implications~\cite{chen_landscape_2024, samuel_strategies_2021}:

\begin{itemize}
    \item Democratization of HPC: By lowering the entry barrier for parallel programming, LLMs have the potential to make HPC more accessible across disciplines, enabling broader participation from researchers without extensive programming backgrounds.
    \item Transformation of skill requirements: As LLMs take on increasingly technical tasks, the demand for deep domain expertise in parallel programming may shift. While this may enhance productivity, it could also alter traditional HPC career paths.
    \item Equity in access: The benefits of LLM-HPC integration may disproportionately favor well-funded institutions with access to high-performance infrastructure and premium models, intensifying existing digital and research inequalities.
\end{itemize}

From an environmental standpoint, the intersection of LLMs and HPC amplifies energy consumption concerns~\cite{chen_landscape_2024, peng_large_2024}:

\begin{itemize}
    \item Training costs: The development of state-of-the-art LLMs requires substantial computational resources, contributing significantly to energy usage and associated carbon emissions.
    \item Inference overhead: Deploying LLMs in HPC environments introduces additional runtime and memory demands, especially at scale, which may offset some of the efficiency gains they offer.
    \item Paradoxically, LLMs may also contribute to energy efficiency by recommending more sustainable algorithm choices, optimizing resource scheduling or reducing unnecessary computation through smarter code generation.
\end{itemize}

\reviewSI{
\section*{Acknowledgments}
We thank the anonymous reviewers for their insightful comments, particularly for encouraging the update of the most recent contributions from late 2025, which has significantly improved the quality of this paper.
}

\section*{CRediT authorship contribution statement}
\begin{itemize}
    \item \textbf{Strahinja Ljaljevic:} Methodology, Investigation, Writing - Original Draft. 
    \item \textbf{Josep Jorba:} Investigation, Writing - Review \& Editing. 
    \item \textbf{Sergio Iserte:} Conceptualization, Investigation, Writing - Review \& Editing, Supervision.
\end{itemize}


\section*{Declaration of competing interest}
The authors have no conflicts of interest to declare that are relevant to the content of this article.

\section*{Funding}
The authors did not receive any specific grant from funding agencies in the public, commercial, or not-for-profit sectors for this research.

\section*{Data availability}\label{sec:data}
\reviewSI{The datasets, figures, and bibliographic references supporting this study are openly available in a dedicated GitHub repository: \url{https://github.com/sljaljevic/llm-hpc-survey}. The repository includes the metadata of the 32 reviewed research papers, the figures used in the manuscript, and the spreadsheets used for data collection and categorization. Researchers are encouraged to reuse and extend these resources in future studies.}

\section*{Declaration of generative AI and AI-assisted technologies in the writing process}
Language polishing was performed using an AI language model, with subsequent thorough review and editing by the authors. The authors take full responsibility for the final manuscript content.

\bibliographystyle{unsrt}
\bibliography{bib/bib, bib/review1}

\end{document}